\documentclass{article}
\usepackage{ amsfonts}
\newcommand{\A}{{\cal A}}
\newcommand{\B}{{\cal B}}
\newcommand{\CH}{{\cal H}}
\newcommand{\CS}{{\cal S}}
\newcommand{\tr}{\mathbf{tr}}
\newcommand{\qed}{\penalty10000\hfill$\Box$\par\goodbreak\medskip}

\newcommand{\definition}{${}$\\[0.3 cm]\noindent{\bf Definition.} } 

\newtheorem{Lemma}{Lemma}
\newtheorem{Theorem}[Lemma]{Theorem}
\begin{document}

\title{Classical Coding and the Cauchy-Schwarz Inequality}
\author{$^*$Bas Janssens}
\date{October 2006}
\maketitle
\begin{center} 
$^*$ Mathematical Institute, Utrecht University 
\end{center}

\abstract{In classical coding, a single quantum state is encoded into 
classical information. Decoding this classical
information in order to regain the original quantum state is
known to be impossible.
However, one can attempt to construct a state which comes as close as 
possible.
We give bounds on the smallest possible trace distance 
between the original and the decoded state
which can be reached.
We give two approaches to the problem: one starting from Keyl and Werner's
no-cloning theorem \cite{KW}, and one starting from an 
operator-valued Cauchy-Schwarz inequality. 
}

\section{Introduction}
There exist results which are possible in 
classical physics, yet  forbidden 
in quantum mechanics.
For instance, it is generally impossible to perform a 
joint measurement on two observables, or to clone one 
quantum state to two identical ones.
This yields a class of problems:
exactly how close can one approximate these
desired, yet impossible results?

We examine this problem in the specific case of 
`classical coding'. Impossibility of
classical coding is the following statement:
if a single quantum state is encoded into classical information,
then it is impossible, starting from this 
classical information, to reconstruct the original quantum state.

In order to 
quantify how far one is removed 
from the ideal situation, 
we define
$\Delta$ to be
the worst case trace distance between the original and the 
decoded state. The impossibility of classical coding
is then equivalent to $\Delta \neq 0$.
The aim is now to investigate which values of $\Delta$
can be reached. We will do this in two different ways.
\begin{itemize}   
\item[-]
The heart of this paper is 
the operator-valued Cauchy-Schwarz inequality,
described in section \ref{seccs}. 
We will use it to obtain the bound 
$\Delta \geq (3 - \sqrt{5})/4$.
\item[-]
We will also attack the same problem using a
no-cloning theorem \cite{KW}, which will yield
the superior bound $\Delta \geq 1/3$.
\end{itemize}
The Cauchy-Schwarz method  
was used earlier in \cite{ik} for the three quantum impossibilities
called
`no measurement without disturbance', `no measurement
without decoherence' and `no joint measurement'. 
In contrast with the case considered here, the 
bounds in these three cases did turn out to be sharp. 
The purpose of this paper is mainly to illustrate the flexibility of 
the Cauchy-Schwarz method.
\newpage
The article is structured as follows:
in section \ref{qm}, a short introduction to quantum probability theory is
given. In section \ref{klaskonijn}, the problem of classical coding
is formulated in a rigorous mathematical fashion. 
In section \ref{seccs}, the Cauchy-Schwarz inequality is given,
and applied to the problem of classical coding. This yields the bound
$\Delta \geq (3 - \sqrt{5})/4$. 
Section \ref{secclo} provides the superior bound $\Delta \geq 1/3$, based on the no-cloning 
theorem. A short discussion of the results then follows in section \ref{disco}.

\section{Quantum Mechanics}\label{qm}

A quantummechanical system is described by a von Neumann algebra $\A$
of bounded operators on a Hilbert space $\CH$,  
usually the algebra $B(\CH)$ of all bounded operators.
Its state space is formed
by the normalized density matrices $\CS (\A ) = \{ \rho \in
\A  \, ; \, \rho \geq 0, \, \tr ( \rho ) = 1\}$. 
With the system in state $\rho \in \CS (\A)$,
observation of a (Hermitean) observable $A \in \A$ 
is postulated to yield the average value $\tr(\rho A)$.  

\subsection{Completely Positive maps}\label{trans}
The natural notion of a map between von Neumann-algebras
is that of a completely positive (or CP for short) map.
\definition Let $\A$ and $\B$ be von Neumann algebras.
A weakly continuous map $T : \B \rightarrow \A$ is called 
\emph{Completely Positive}
(or CP for short) if it is linear, normalized 
(i.e. $T(I) = I$), positive (i.e. $T(X^{\dagger}X) \geq 0$ for all $X \in \B$) and
if moreover the extension 
${\it Id}_{n} \otimes T : M_n \otimes \B \rightarrow M_n \otimes \A$
is positive for all $n \in {\mathbb{N}}$, where $M_n$ is the algebra of
complex $n \times n$-matrices. 
\\[0.3 cm]
\noindent Its dual $T^{*} : \CS(\A) \rightarrow \CS(\B)$, defined by the
requirement $\tr(T^*(\rho) X) = \tr(\rho T(X))$ $\forall \, X \in \B$,
has a direct
physical interpretation as an operation between quantum systems.
$T$ is positive, linear and normalized. This is equivalent 
to $T^*$ being an affine map $\CS(\A) \rightarrow \CS(\B)$.
That is, each state $\rho
\in \CS(\A)$ is again mapped to a state $T^* (\rho) \in \CS (\B)$, and 
for all $p \in [0,1]$ and $\rho_1 , \rho_2 \in \CS(\A)$, we have
 $p T^* (\rho_1) + (1-p) T^*(\rho_2) =  T^* (p \rho_1
+ (1-p) \rho_2)$. 
This expresses the 
\emph{stochastic equivalence principle}: a system which is in state $\rho_1$
with probability $p$ and in state $\rho_2$ with probability $(1-p)$ cannot
be distinguished from a system in state  $p \rho_1 + (1-p) \rho_2$.

It is possible to extend the systems $\A$ and $\B$
under consideration with another system $M_n$, on which the operation
acts trivially.  
Due to \emph{complete} positivity, states in $\CS (M_n \otimes \A)$ are once again 
mapped to states in $\CS(M_n \otimes \B)$.
This is of course a crucial
property if one seeks to interpret $T^*$ as a physical map: 
the mere act of coupling the system $\A$ to another system 
$M_n$ may never cause negative probabilities.   
Surprisingly enough, there exist linear maps which are positive,
but not completely positive.
It is generally believed that {\it any} operation which can be 
physically implemented on a quantum system is described by a CP-map.

\subsection{Probability Spaces}

We are interested in quantum operations which take as input a 
quantum state, and yield a 
classical probability distribution as output. 

In order to encompass this in our framework of CP-Maps
and von Neumann algebras, 
we identify a classical probability space with a
commutative von Neumann algebra. (See \cite{Hans2}.)
A classical probability space $(\Omega , \Sigma , \mathbb{P})$ 
gives rise to $\A := L^{\infty}(\Omega , \Sigma , \mathbb{P})$,
the set of bounded measurable functions on $\Omega$
up to the equivalence $\sim$,
with $f \sim g$ if $f = g$ almost surely.
We think of these as the {\it random variables}.
Since $f \in \A$ acts on the Hilbert space $L^2(\Omega)$
by multiplication, we can regard $\A$ as a 
commutative subalgebra of $B(L^2(\Omega))$, which 
turns out to be weakly closed. The probability measure
$\mathbb{P}$ of course induces the expectation 
$\mathbb{E}(f) = \int_{\Omega} f(\omega) \mathbb{P}(d\omega)$
of a random variable. This yields a state $\rho_{\mathbb{P}}$
by requiring $\tr( \rho_{\mathbb{P}} M_f) := \mathbb{E}(f)$.

In short, a classical probability space corresponds to a 
commutative von Neumann algebra, and a probability measure 
corresponds to a state on that algebra.
An operation $C^*$ which maps quantum states in $\CS(B(\CH))$ to 
classical probability distributions in $\CS(\A)$ must therefore
be the dual of a CP-map $C : \A \rightarrow B(\CH)$.

For example, the direct measurement of an observable (= bounded 
Hermitean operator) $X$. 
With each Hermitean operator, there is associated a projection 
valued measure $P(dx)$, such that 
$X = \int_{\mathbf{Spec}(X)} x P(d x)$. 
The measurement of $X$ 
is then represented by the 
CP-map $L^{\infty}(\mathbf{Spec}(X)) \rightarrow B(\CH)$ defined by
$f \mapsto \int_{\mathbf{Spec}(X)} f(x) P(dx)$.
In the dual (Schr\"odinger) picture, we then see that a state 
$\rho \in \CS(B(\CH))$ induces the probability distribution
$\mathbb{P}(dx) = \tr( \rho P(dx) )$ on the spectrum of
$X$. More generally, each CP-map 
$L^{\infty}(\Omega) \rightarrow B(\CH)$ corresponds 
to a Positive Operator Valued Measure, but we will not 
need this here. 

Similarly, an operation $D^*$ which maps classical probability distributions
into quantum states must be the dual of a CP-map 
$D : B(\CH) \rightarrow \A$.

\section{Classical Coding}\label{klaskonijn}

By \emph{classical coding}, we mean the following procedure.
First, classical information is extracted from a quantum system.
This is described by the coding map $C : \A \rightarrow B(\CH)$,
with $\A = L^{\infty}(\Omega)$ for some space $\Omega$.
Then, on the basis of this classical information, 
the original state is reconstructed as well as possible 
by means of some decoding procedure. This is described by a CP-map  
$D : B(\CH) \rightarrow \A$. In the dual picture, this gives
$$
\CS(B(\CH)) \stackrel{C^*}{\rightarrow} \CS(\A) 
\stackrel{D^*}{\rightarrow} \CS(B(\CH))\,.
$$
The coding procedure is flawless iff every state is reconstructed
perfectly, i.e. iff $C\circ D : B(\CH) \rightarrow B(\CH)$ is the identity.  
 
It is well known \cite{We, Hans} that perfect classical coding is impossible.
So let us investigate how close we can come to perfection.
Define $\Delta$ to be the maximum difference 
between input and output probability on a single event $P$, i.e.
$$
\Delta := \sup_{\rho , P} 
|\tr(D^*C^*(\rho) P) - \tr(\rho P) |,
$$  
where $\rho$ runs over $\CS(B(\CH))$, and $P$ runs over the
projections in $B(\CH)$.
We give two reformulations of this definition (see\cite{NC}).
The first is
$\Delta = \sup_{\rho} |D(\rho, D^*C^*(\rho))|$,
with $D(\rho,\tau)$ the
trace distance or Kolmogorov distance 
$D(\rho,\tau) = \frac{1}{2} \tr(|\rho - \tau|)$ between $\rho$ and $\tau$.
The second is $\Delta = \sup_{B} \|B - CD(B)\|$, where $B$ runs
over the positive operators $0 \leq B \leq I$. 
The latter is most convenient, and we will use it in the remainder 
of the article.

We remark that $\Delta$ quantifies the quality of the
coding procedure:
a large value of $\Delta$ corresponds to a poor classical 
coding procedure, a small value of $\Delta$ corresponds to a good one.
Furthermore, $\Delta = 0$ if and only if all states are encoded
perfectly. We now investigate how close to zero $\Delta$ can come. 

\section{The Cauchy-Schwarz Method}\label{seccs}

We start with a lemma which, in all its simplicity, is the 
cornerstone of a veritable zoo of quantum inequalities.
(See \cite{ik,scr}. The special case $(A,A)=0 \Rightarrow
(A,B) =0$ is older, and due to Werner \cite{We}.) 
\begin{Lemma} [Cauchy-Schwarz] \label{CS}
Let $\A$, $\B$ be von Neumann algebras, and let
$(\,\cdot \, ,\, \cdot \, ) : \, \, \cal{A} \times 
\cal{A} \rightarrow \cal{B}$ be
a positive semidefinite sesquilinear form. That is,
it is linear in the second argument, 
$(A,B)^\dagger = (B,A)$ for all $A,B \in \A$, and 
$(A,A) \geq 0$ for all $A \in \A$.
Then 
$\|\Re (A,B)\|^2 \leq \|(A,A)\| \|(B,B)\|$
and
$\|\Im (A,B)\|^2 \leq \|(A,A)\| \|(B,B)\|$
for all $A,B \in \A$.
\end{Lemma}
{\bf Proof}\\*
For $A,B \in \A$ and $\lambda \in \mathbb{C}$, we have 
(with $\Re X := (X + X^\dagger ) /2$ the `real' 
and $\Im X := (X - X^\dagger)/2i $ the `imaginary' part):
\begin{equation}\label{smurf}
0 \leq 
(A - \lambda B , A - \lambda B) =
(A,A) -2\Re \lambda(A,B) + |\lambda|^2 (B,B).
\end{equation}
If $(A,A) = 0$ and $(B,B) = 0$, the lemma follows
immediately from (\ref{smurf}).
If not, assume that $(B,B) \neq 0$, exchanging 
the roles of $A$ and $B$ if necessary.
Choose $\lambda = \pm \|\Re(A,B)\| / \|(B,B)\|$, so that
(\ref{smurf}) becomes the operator inequality
$$
\pm 2 \Re (A,B) \|\Re (A,B)\|/\|(B,B)\| \leq
(A,A) + (B,B) \|\Re (A,B)\|^2 /\|(B,B)\|^2 .
$$
In particular, 
the spectrum of the operator on the l.h.s. 
is contained in $[-\|R\| , \|R\|]$, with
$R$ the r.h.s. operator.
Thus
$
 2 \|\Re (A,B)\|^2/\|(B,B)\| \leq \|R\|$ ,
and $ \|R\| \leq \|(A,A)\| + \|\Re (A,B)\|^2 /\|(B,B)\|
$. This yields $\| \Re(A,B) \|^2 \leq 
\|(A,A)\| \|(B,B)\|$, as 
required. 
Since $\Re (i A,B) = \Im (A,B)$ holds, we also have
$\|\Im (A,B)\|^2 $ $\leq$ $\|(A,A)\| \|(B,B)\|$.
\qed

For example, each CP-map $T : \A \rightarrow \B$
induces a positive semidefinite ses\-quilinear form
by $(A,B)_T := T(A^\dagger B) - T(A)^\dagger T(B)$.
Indeed, according to Stinespring's theorem (see \cite{St}), we can 
assume without loss of generality that $T$
is of the form $T(A) = V^\dagger A V$, with 
$\|V\| \leq 1$.
Then $(A,A)_T = V^\dagger A^{\dagger}(I - VV^\dagger) AV =
(\sqrt{(I - VV^\dagger)} AV)^\dagger 
\sqrt{(I - VV^\dagger)} AV\geq 0$.  
The sesquilinearity is clear.

We use the Cauchy-Schwarz inequality to obtain a
the bound on $\Delta$ for any classical coding procedure.
(The proof is inspired by the `impossibility of classical coding' version
in \cite{Hans}.)
\begin{Theorem}
Let $\A$ be a commutative von Neumann-algebra, 
and let $B(\cal{H})$ be the algebra of bounded operators on a Hilbert space
$\cal{H}$ of dimension $>1$.
Let 
$C : \A \rightarrow B(\cal{H})$ and
$D : \B(\cal{H}) \rightarrow A$ be
CP-maps. 
Let $\Delta := 
\sup \{ \| B - C \circ D (B) \| : 0 \leq B \leq I \}$.
Then $\Delta \geq (3 - \sqrt{5})/4$.
\end{Theorem} 
{\bf Proof}\\*
Take two orthogonal vectors $\psi, \phi \in \cal{H}$,
and define $X$ to be the projection on $\psi$,
and $Y$ the projection on $(\psi + \phi)/\sqrt{2}$.
We have $\|[X,Y]\| = 1/2$. Since $\A$ is Abelian, we have
$D(X)D(Y) = D(Y)D(X)$, and we can write
\begin{eqnarray}\label{mandofiraat}
[X,Y] &=& [X,Y] - CD([X,Y]) + \label{mangtor} \\
  & &  C\left( D(XY) - D(X)D(Y) \right) - \label{koptig}\\
     & & C \left( D(YX) - D(Y)D(X) \right)\,. \nonumber
\end{eqnarray}
We will bound the r.h.s. in terms of $\Delta$. Remembering 
that the l.h.s. is at least $1/2$ in norm will then yield
a minimum value on $\Delta$.  

We start with (\ref{mangtor}).
Like any antihermitean operator, $[X,Y]$ can be
written as $[X,Y] = i(A_+ - A_-)$, with 
$0 \leq A_\pm \leq \|[X,Y]\| I = \frac{1}{2} I$. Therefore,
we have $\|[X,Y] - CD([X,Y])\| \leq 
\Delta \|A_+\| + \Delta \|A_-\| \leq \Delta$.

We then proceed with (\ref{koptig}).
Consider the positive semidefinite sesquilinear form
$(X,Y) := C\left( D(X^\dagger Y) - D(X)^\dagger D(Y)\right)$,
in terms of which the expression (\ref{koptig}) equals
$2i\Im (X,Y)$.
It is positive as the concatenation of 
$( \,\cdot \, , \, \cdot \,)_D : 
B(\cal{H}) \times B(\cal{H}) \rightarrow \A$ and
the positive map $C : \A \rightarrow B(\cal{H})$.

According to Lemma \ref{CS}, we have
$\| 2i\Im (X,Y)\| \leq 2 \sqrt{\|(X,X)\|\|(Y,Y)\|}$.
Now $(X,X) = (X,X)_{C D} -(D(X),D(X))_C \leq 
(X,X)_{C D}$, and 
a similar expression holds for $Y$.

Since $X^\dagger X = X$, we have
$(X,X)_{C D} = C D (X) (I - C D (X))$.
Since $\|X - CD(X)\| \leq \Delta$, and
$X$ has spectrum $\{0,1\}$,  
the spectrum of $CD(X)$ is contained in 
$[0,\Delta] \cup [1-\Delta,1]$.
(Recall that $0 \leq CD(X) \leq I $.)
The spectrum of $C D (X) (I - C D (X))$
therefore lies within $[0,\Delta(1-\Delta)]$,
so that $\|(X,X)_{C D}\| \leq \Delta(1-\Delta)$.
The same holds for $Y$ of course.
Putting this together, we bound (\ref{koptig})
as $\| 2i\Im (X,Y)\| \leq 2 \Delta(1-\Delta)$.  

We conclude that $1/2 = \|[X,Y]\| \leq \Delta + 2\Delta(1-\Delta)$,
or $(\Delta - 3/4)^2 \leq 5/16$. Thus 
$\Delta \geq (3 - \sqrt{5})/4$, which was to be proven. \qed

\section{The No-Cloning method}\label{secclo}
It is easy to obtain a better bound from the no-cloning theorem
of Keyl and Werner \cite{KW}.
The idea is, that each classical coding 
procedure yields a cloning machine.
All one has to do is to `decode' the classical information 
an arbitrary amount $M$ of times, rather than just once.

To be more explicit,
suppose that we are in the finite setting:
$\CH = \mathbb{C}^{d}$ is a finite-dimensional Hilbert space, and 
$\A = L^{\infty}(\Omega)$  with  $\Omega = \{1,2, \ldots ,n\}$.
The `diagonal' map $\Omega \rightarrow \Omega^{M} \,:\, 
i \mapsto (i, \ldots ,i)$ induces the `classical 
cloning' map $K : \A^{\otimes M} \rightarrow \A$, i.e.
$(K f)(i) := f(i,\ldots,i)$. 
(Note that for e.g. $\Omega = \mathbb{R}$, a cloning map 
poses difficulties.)

The composition $ T := C \circ K \circ D^{\otimes M}$, mapping
$B(\mathbb{C}^{d})^{\otimes M}$ to $B(\mathbb{C}^{d})$,
is a so-called $1 \rightarrow M$ cloner.
By construction of $T$, it is clear that
$T(I \otimes \ldots \otimes B \otimes \ldots \otimes I)
=CD(B)$.
The main result of \cite{KW} then says that 
$\sup_{0 \leq B \leq I} 
{\|T(I \otimes \ldots \otimes B \otimes \ldots \otimes I) - B\|}
\geq \frac{(M-1)}{M}\frac{d-1}{d+1}$.
Since $M \in \mathbb{N}^+$ was arbitrary,
this implies $\Delta \geq \frac{d-1}{d+1}$.
The best possible coding occurs for $d=2$, when 
$\Delta \geq \frac{1}{3}$.

\section{Discussion} \label{disco}

The number $(3 - \sqrt{5})/4 \approx 0.19$, obtained from the
Cauchy-Schwarz inequality, is inferior to the $\frac{1}{3} \approx 0.33$
from the no-cloning theorem, in usefulness as well as in
the standard order on $\mathbb{R}$.  

Although it yields inferior results for classical coding, 
the Cauchy-Schwarz method  is simpler. At least taking into account
the fairly heavy machinery needed in \cite{KW}.
It is also very flexible.
For any kind of quantum impossibility, ranging from
`no classical coding' to `no measurement without disturbance',
and from `decoherence after measurement' to 
`no joint measurement', 
the Cauchy-Schwarz inequality
yields a bound which quantifies how far one is 
removed from the ideal, impossible situation. (See \cite{ik,scr}.)
   
Surprisingly enough, in the 
cases of `no measurement without decoherence'
`no measurement without disturbance' and
`no joint measurement', 
this bound becomes sharp in the sense that 
there exist quantum operations which reach it. 
Furthermore, as far as I am aware, 
these three bounds cannot be obtained in any other way.

In conclusion, the Cauchy-Schwarz method yields 
quantitative bound for a variety of 
quantum impossibilities, including classical coding.
Although this is not the case for classical coding,
many of these bounds turn out to be sharp.


\end{document}